# Predicting biomolecular binding kinetics: A review

Jinan Wang[1], Hung N. Do[1], Kushal Koirala[1], Yinglong Miao[1,*]

[1]Center for Computational Biology and Department of Molecular Biosciences, University of Kansas, Lawrence, Kansas 66047

*To whom correspondence should be addressed: miao@ku.edu

**Abstract**

Biomolecular binding kinetics including the association ($k_{on}$) and dissociation ($k_{off}$) rates are critical parameters for therapeutic design of small-molecule drugs, peptides and antibodies. Notably, drug molecule residence time or dissociation rate has been shown to correlate with their efficacies better than binding affinities. A wide range of modeling approaches including quantitative structure-kinetic relationship models, Molecular Dynamics simulations, enhanced sampling and Machine Learning have been developed to explore biomolecular binding and dissociation mechanisms and predict binding kinetic rates. Here, we review recent advances in computational modeling of biomolecular binding kinetics, with an outlook for future improvements.

## 1. Introduction

Life processes are critically dependent on the formation of biomolecular complexes, particularly including the protein-small molecule, protein-peptide and protein-protein/antibody structures. Biomolecular binding plays a key role in many fundamental biological processes[1]. Accurate characterization of biomolecular binding thermodynamics and kinetics is key for therapeutic design[2]. Recently, drug residence time or dissociation rate appears to correlate with drug efficacy better than the binding free energy[3]. With remarkable theoretical and technical developments, increasing numbers of experimental and computational methods are available for calculating the biomolecular binding kinetic rates[3a, 3e, 3h, 4]. However, it remains challenging for both experimental and computational approaches to accurately predict biomolecular binding kinetic rates with high throughput.

In this review, we will first briefly describe available experimental techniques for determining biomolecular binding kinetic rates. We will then discuss computational approaches to predict the biomolecular binding kinetics, with focus on the Molecular Dynamics (MD) and enhanced sampling methods, which have emerged as rapidly evolving techniques for studying biomolecular binding kinetics.

## 2. Available experimental techniques to measure binding kinetics

Most experimental techniques[5] for determining biomolecular binding kinetic rates are mainly relying on monitoring a specific signal over time during the binding and dissociation processes. According to signal source, experimental methods could be generally divided into two classes: assays with and without a label for detection[4b]. Radio and spectroscopic labeling are the main choices for labeling assays. A radiolabel essentially comes from the presence of radioactive isotopes in the molecule, which could emit special radiation when they decay to more stable states. In radiometric binding assays, ligands are tagged to follow the time course of their binding to targets, thus allowing for the spontaneous measurement of binding kinetic rates[6]. In the spectroscopy-based assays, ligands are labeled with fluorophore groups. After absorbing a certain wavelength's light, fluorophore groups could emit characteristic light, allowing for detecting the binding and dissociation processes[7]. The fluorescent resonance energy transfer (FRET) is one popular spectroscopy based approach[7]. For the label free approaches, surface plasmon resonance (SPR) is one of the most widely used methods, especially in characterizing the biomolecular binding kinetics of pharmaceutical interest[2a].

With developments of experimental techniques, recent years have seen significantly increasing numbers of biomolecular binding kinetic data, including the protein-small molecule, protein-peptide and protein-protein binding kinetic rate constants. Many experimental binding kinetic rates have been collected in different publicly accessible databases. A number of databases as listed in **Table 1** are useful for exploring biomolecular binding kinetics, including the kinetic data of bio-molecular interactions (KDBI)[8], BindingDB[9], kinetics of featured interactions (KOFFI)[10], PDBbind[11], structural database of kinetics and energetics of mutant protein interactions (SKEMPI)[12], kinetic and thermodynamic database of mutant protein interactions (dbMPIKT)[13] and so on[3d, 14].

KDBI[8] is developed to provide experimentally verified binding kinetic rates for interactions involving proteins and nucleic acids (RNA and DNA). It includes 19,263 entries of 10,532 distinguishing biomolecular pathways. The binding kinetic data includes protein-protein/DNA/RNA/ligand and ligand-DNA/RNA interactions. BindingDB[9] is one widely used database for exploring protein-small molecule interactions, containing ~1.1 million compounds and 8.9 thousand targets with clearly defined quantitative measurement for binding affinities and kinetic rates. BindingDB provides a special kinetic database via link https://bindingdb.org/rwd/bind/ByKI.jsp?specified=Kn. The data of BindingDB are extracted from published literatures and other databases such as PubChem, CheEMBL, PDSP Ki, and CSAR. Additionally, BindingDB provides an option for experimentalists to directly deposit their data. KOFFI[10] is developed to provide binding kinetic rates along with experimental protocol. It includes 1705 individual entries. Notably, it contains a rating system to assess quality of experimental data. A user can perform a direct search within the Anabel's KOFFI database and evaluate the quality of their binding data. PDBbind[11] was initially developed for collecting binding affinity data and complex structures for developing docking score. In 2022, it released a sub-database ($k_{off}$ set) containing 169 entries of protein-small molecule dissociation rates. One advantage of PDBbind is the availability of the protein-small molecule complex structures, which could be convenient for molecular modeling. SKEMPI[12] and dbMPIKT[13] mainly focus on protein-protein interaction (PPI). SKEMPI[12] contains 713 protein–protein binding kinetic rates upon mutation. dbMPIKT[13] contains 5291 entries of protein-protein binding kinetic rates involving mutation. In summary, developments of experimental techniques and increasing biomolecular binding kinetic data collected in the databases will greatly facilitate modeling of biomolecular binding kinetics and therapeutic design.

## 3. Quantitative structure-kinetic relationships

Optimal kinetic parameters for biomolecular binding could significantly improve drug efficacy. For that reason, several molecular modeling techniques have been developed to predict biomolecular binding kinetic rates and derive quantitative structure-kinetic relationships (QSKRs)[15]. While these methods are often based on experimental structures, many of them consider each biomolecular complex with only one single structure[15]. Nunes-Alves et al.[15] modified the COMparative BINding Energy (COMBINE) analysis, which uses *holo* structure to predict binding parameters, to include extra options of using multiple protein-small molecule complex structures. They did so by docking small molecules to a protein conformational ensemble obtained from MD simulations. Specifically, full data set for COMBINE analysis consisted of 33 inhibitors of p38 MAP kinase, which were chosen given availability of experimental $k_{off}$ values and experimental structures of the inhibitor bound to p38 MAP kinase or to other kinases in the DFG-out conformation state. 22 and 11 inhibitors were used for training and testing in the COMBINE analysis, respectively. The first step in the COMBINE analysis involved modelling of the two sets of structures and derivation of COMBINE analysis models. After energy minimization of the complex structures, interaction energy components were obtained with the AMBER ff14SB force field to describe bonded and non-bonded interactions. Weights to scale the protein-small molecule interaction energies were obtained using partial least square regression. To account for multiple structures, the COMBINE was modified to retrieve an average response using N structures for each protein-small molecule complex, in which each structure was treated independently during regression to obtain weights for interaction energies. Here, exponential or arithmetic averages could be used:

$$(\log I)_{exp}^{comp} = -log\frac{1}{N}\sum_{j=1}^{N} e^{-logI^j} \tag{1}$$

$$(\log I)_{arit}^{comp} = \frac{1}{N}\sum_{j=1}^{N} log\ I^j \tag{2}$$

where $(\log I)_{exp}^{comp}$ and $(\log I)_{arit}^{comp}$ were the predictions for the response variable using exponential and arithmetic averages, $j$ was the index of the structure used, $log\, I^j$ was the prediction made using the $j^{th}$ structure, and $N$ was the number of structures to describe one protein-small molecule complex. In one of the two structure sets used for the COMBINE analysis, each complex was represented using one experimental crystal structure. In another set, each complex was represented using 10 structures from ensemble docking[15]. Although COMBINE model obtained with multiple structures from ensemble docking took protein-ligand flexibility into consideration, the predictive power was lower than the model from a single, energy-minimized crystal structures for each protein-ligand complex. Nevertheless, the incorporation of protein-ligand flexibility highlighted additional important protein-ligand interactions that led to longer residence time.

In another study, Schuetz et al.[16] performed matched molecular pair (MMP) analysis on datasets assembled from the Kinetic for Drug Discovery database, which included 3812 small molecules annotated to 78 different targets from five diverse protein classes, including G-protein-coupled receptors (GPCRs), kinases and other enzymes, heat shock proteins (HSPs), and ion channels. The kinetic dataset (KIND) contains complex structures along with their respective binding kinetic data ($k_{on}$, $k_{off}$, $K_D$). To elucidate the impact of small structural changes on the binding kinetic behavior, a total of 395 MMPs extracted from KIND were performed. The pairs were made of two molecules possessing identical scaffolds and showing minor chemical modifications. This dataset included various chemical modifications, with the top 20 representing less than 65% of the entire dataset. The most common modification, which was replacement of a hydrogen atom by a methyl group, comprised around 15%. To demonstrate that changes in a molecule's polarity are the major factor for the alteration of binding association rate $k_{on}$, the authors focused on analyzing top 20 MMPs with highest differences in $k_{on}$ values. For 16 out of 20 MMPs, a substitution that increases polarity was observed. The largest differences in $k_{on}$ were found with the introduction of charged moiety, leading to decrease of 0.5- 2.0 orders of magnitude. The decrease in $k_{on}$ might come from electrostatic repulsion and/or desolvation penalties. Conversely, an improvement in binding affinity was observed if modifications established additional interactions in the final bound complexes. The dissociation rate $k_{off}$ was also analyzed following the same protocol for $k_{on}$. In contrast to $k_{on}$, the change of molecular polarity in the MMPs did not produce a consistent shift in $k_{off}$.

In 2018, Ganotra and Wade applied COMBINE analysis to derive QSKRs for the dissociation rates ($k_{off}$) of inhibitors of HSP90 and HIV-1 protease[17]. Protein-specific scoring functions were derived by correlating $k_{off}$ with a subset of weighted interaction energy components determined from energy minimized biomolecular complex structures. A set of 3D structures of protein-ligand complexes were modeled and energy minimized. Protein-ligand interaction energies were first calculated, then partitioned and subjected to partial least-squares projection to latent structures (PLS) regression. A statistical model was derived to correlate the activity of interest to weighted selected components of the protein-ligand interaction energy decomposed on a per residue basis, based on the following equation:

$$\log(k) = \sum_{i=1}^{n} w_i \Delta u_i + C \tag{3}$$

where $k$ was the rate constant of interest, and $\Delta u_i$ were per residue terms of the ligand-receptor interaction energy, calculated for $n$ residues. The coefficients $w_i$ and constant $C$ could be determined from PLS regression. The dataset used for the COMBINE analysis of HSP90 and HIV-1 protease inhibitors consisted of 70 and 36 compounds, respectively. Experimental $k_{off}$ values ranged from 0.0001 to 0.83 $s^{-1}$ for the HSP90 inhibitors and 0.00022 to 83.3 $s^{-1}$ for the HIV-1 protease inhibitors. For the COMBINE analysis, 207 coulombic and 207 Lennard-Jones (LJ) interaction energy terms were calculated for the HSP90 inhibitors, and 198 coulombic and 198 LJ energies were calculated for HIV-1 protease inhibitors. The resulting COMBINE models for $k_{\text{off}}$ rates had very good predictive power ($Q^2_{\text{LOO}}$ = 0.69 for HSP90, and $Q^2_{\text{LOO}}$ = 0.70 for HIV-1 protease), which could also identify contributing protein-ligand interactions for binding kinetics.

In order to explore molecular details of biomolecular binding processes on a large scale, Chiu et al. [18] recently integrated coarse-grained normal mode analysis (NMA) with multi-target machine learning (MTML) to address the above challenge and tested their method using the HIV-1 protease as a model system. The workflow included four phases. In phase 1, the 3D complex structure of the ligand-bound HIV-1 protease was built. Ligands without experimental structure were docked into the HIV-1 protease using the eHiTS software. In phase 2, residues in the ligand-binding site were identified. Coarse-grained NMA was performed for both *apo* and *holo* structures. The authors defined RMLR as the dot product of ligand displacement vector after normalization and the residue displacement vector, and RMRR as the dot product of the displacement vectors of a residue for the *apo* and *holo* structures. Therefore, RMLR and RMRR could be derived from the NMA and describe the conformational dynamics impact of ligand binding on the residues in the binding site. In phase 3, five principal data sets were constructed. Pairwise decomposition of the residue interaction energy was computed by minimizing 39 ligand-bound HIV-1 complexes with NAMD simulations using the generalized Born implicit solvent (GBIS) method. The final simulation conformations were used to compute the residue-decomposed pairwise interaction energy (PIE), the van der Waals energy (VDWE) and the electrostatic energy (EE) between the ligand and protein residues. The energetic features (PIE, VDWE, and EE) and conformational dynamics features (RMRR and RMLR), along with experimentally determined $k_{\text{on}}$ and $k_{\text{off}}$ data were used to train MTML models in phase 4 of the workflow. The model was evaluated regarding the accuracies in the predictions of binding kinetic rate constants $k_{\text{on}}$ and $k_{\text{off}}$ using the following formula:

$$\text{accuracy} = \sum_{i=1}^{n} \frac{A_i}{N} \tag{4}$$

where $A_i$ was the prediction accuracy for each case and $N$ was the total number of cases. $A_i = 100\%$ when both $k_{\text{on}}$ and $k_{\text{off}}$ were accurately predicted, and $A_i = 0\%$ when neither was correctly predicted. The model was further evaluated in high-throughput screening of molecules with *in vivo* drug activity on the basis of $k_{\text{on}}$ and $k_{\text{off}}$ using the receiver operating characteristic (ROC) curve and the area under the ROC curve (AUC). The computational models were not only found to recapitulate the results from MD simulations but also accurately predict protein-ligand binding kinetic rates, with an accuracy of 74.35% when combined with energy features. In addition, the integrated models showed that the coherent coupling of conformational dynamics and thermodynamic interactions between the receptor and ligand played a critical role in determining protein-ligand binding kinetic rates.

Engel et al. designed novel and irreversible epidermal growth factor receptor (EGFR) inhibitors using a structure-based approach rationalized by subsequent computational analysis of conformational ligand ensembles in solution[19]. The approach was based on a screening hit that was identified in a phenotype screen of ~1,500 compounds in 80 non-small cell lung cancer (NSCLC) cell lines. With X-ray crystallography, the binding mode in engineered cSrc (T338M/S345C), a validated model system for EGFR-T790M, was deciphered. Chemical synthesis revealed further compound collections that increased biochemical potency and selectivity toward mutated (L858R and L858R/T790M) vs. wildtype EGFR. Kinetic studies were performed to investigate the rate and efficiency of covalent bond formation for the most effective inhibitors, **5b** and **6a**. The corresponding time-dependent affinity ($K_i$) and reactivity ($k_{inact}$) parameters with respect to the mutants L858R and T790M/L858R of EGFR were determined by an activity-based assay. The respective $IC_{50}$ values of **5b** and **6a** were monitored after treatment of the respective proteins in a time-dependent manner. These values were correlated to the respective incubation times, from which $K_i$ and $k_{inact}$ parameters could be determined[8]. Inhibitors **5b** and **6a** were found to exhibit extraordinarily high affinity toward EGFR L858R/T790M, with respective $K_i$ values of 0.64 and 0.32 nM, and specific, moderate reactivity, with $k_{inact}$ values of 0.116 and 0.137 $min^{-1}$. On the contrary, both the binding affinity and specific reactivity of **5b** and **6a** toward EGFR L858R were significantly impaired ($K_i$ = 70.2 and 833 nM, and $k_{inact}$ = 0.017 and 0.055 $min^{-1}$). In summary, with increasing numbers of available experimental binding kinetic data and advances in the modeling approaches, the built QSKR will become more accurate and allow for high-throughput screening, which is very helpful at early stage of drug design.

## 4. Molecular Dynamics and enhanced sampling methods for predicting binding kinetics

MD is a powerful technique for simulations of biomolecular structural dynamics[20]. The accessible timescale of conventional MD (cMD) has reached hundreds of microseconds thanks to remarkable advances in computing hardware (e.g., the Anton supercomputer and GPUs) and software developments[21]. Notably, the latest Anton3[21f] has enabled hundreds-of-microseconds cMD simulations per day. The cMD simulations have been widely applied to investigate biomolecular binding processes[22]. However, it is still challenging for cMD to simulate repetitive biomolecular dissociation and rebinding processes[21a, 23]. In this regards, enhanced sampling methods[24] have been developed to simulate biomolecular binding and dissociation processes, and predict the associated binding kinetic rates, including the widely used Weighted Ensemble[25], mile-stoning method[26], Gaussian accelerated MD (GaMD)[27], Metadynamics[28], Markov State Modeling (MSM)[29], Random Acceleration Molecular Dynamics (RAMD) [30], scaled MD [31] and so on. Recent years have seen a significant increasing numbers of studies on predicting biomolecular binding kinetic rates using MD simulations (**Fig. 1A**). To evaluate the accuracy of simulation predicted kinetic rates, we define the prediction errors of binding and dissociation kinetic rates as:

$$\Delta \log k_{on} = \log k_{on}^{sim} - \log k_{on}^{exp},$$
$$\Delta \log k_{off} = \log k_{off}^{sim} - \log k_{off}^{exp}, \quad (5)$$

where simulation predicted binding ($k_{on}^{sim}$) and dissociation ($k_{off}^{sim}$) rates are compared with experimentally determined binding ($k_{on}^{exp}$) and dissociation ($k_{off}^{exp}$) rates. Most values of the $\Delta \log k$ are in the range of -1 to 1 (**Fig.1B**), suggesting good prediction accuracy obtained from MD simulations. In the next sections, we will discuss recent applications of the above-mentioned methods in exploring biomolecular binding kinetics for distinct protein-small molecule, protein-peptide and protein-protein binding systems.

**Protein-small molecule binding kinetics**

Compared with slower ligand dissociation process, ligand binding is much quicker, which allows cMD to capture ligand binding process and predict the binding association rate ($k_{on}$). For example, spontaneous binding of the Dasatinib drug to its target Src kinase was observed in a total of ~35.0 μs cMD simulations performed by Shan et al.[22a]. The estimated binding association rate ($k_{on}$) was $0.19 \times 10^7$ $M^{-1}s^{-1}$, being highly consistent with the experimental value of $0.5 \times 10^7$ $M^{-1}s^{-1}$. The same system was used to test a novel approach-unaggregated unbiased MD (UUMD) developed by Sohraby et al.[32]. In contrast to the repulsion added to special atom in the ligand by Shan et al.[22a], the repulsion in the UUMD was added to a virtual interaction site in the ligand to avoid aggregation. Notably, the UUMD could capture multiple independent Dasatinib binding events within nanosecond simulations. The predicted binding association rate ($k_{on}$) was $0.75x10^7$ $M^{-1}s^{-1}$, being highly consistent with the experimental data (**Table 2**). It is worth noting that no dissociation event was observed in the cMD simulations, prohibiting calculation of ligand dissociation rate ($k_{off}$).

Coarse-grained models were developed for MD simulations to reduce the demands for computational resources and extend the simulation timescale[33]. Based on Martini coarse-grained model, Dandekar et al.[34] captured spontaneous binding of benzamidine to the trypsin binding pocket from bulk solvent. Based on 426 μs MD simulation data, they predicted the binding kinetic rates of ($k_{on}$, $k_{off}$) at ($36.8x10^7M^{-1}s^{-1}$, $6.9$ $x10^5s^{-1}$). The corresponding experimental values were ($2.9x10^7M^{-1}s^{-1}$, $600s^{-1}$). Therefore, the predicted $k_{on}$ value was ~13 folds higher than the experimental data. However, large derivation was observed between the predicted and experimentally determined $k_{off}$.

Multiscale computational approaches have been developed to improve the efficiency and accuracy of ligand binding thermodynamics and kinetics calculations[35]. For example, simulation enabled estimation of kinetic rates (SEEKR) [35b, 36] is a multiscale simulation approach combining MD, Brownian dynamics, and mile-stoning for predicting protein−ligand binding association and dissociation rates. SEEKR has been shown to estimate accurate binding kinetic rates with simulation time reduced by a factor of 10[35b]. Using the trypsin-benzamidine model system as example, the SEEKR and its latest version SEEKR2 predicted the binding kinetic rates of ($k_{on}$, $k_{off}$)

at ($12\pm0.5x10^7M^{-1}s^{-1}$, $174\pm9s^{-1}$) and ($2.4\pm0.2x10^7M^{-1}s^{-1}$, $990\pm130s^{-1}$), respectively, being highly consistent with the corresponding experimental data of ($2.9x10^7M^{-1}s^{-1}$, $600s^{-1}$).

Mile-stoning method[26] has been applied to predict the dissociation rate of the Imatinib drug to Abl kinase. Based on the total of 1.043 μs simulations, the value of $k_{off}$ was predicted as 18 $s^{-1}$, being highly consistent with the experimental value of $25\pm6s^{-1}$.Weighted Ensemble[37] and MSM[29a] have been developed to improve prediction of ligand binding kinetic rates based on a large number of short cMD trajectories. In the trypsin-benzamidine system, the dissociation rate ($k_{off}$) of 2,660 $s^{-1}$ was predicted with one weighted ensemble[37] of a total amount of 8.75 μs cMD simulations, being ~4.43 times faster than the experimental value. Another weighted ensemble[38] of a total of 0.48 μs cMD was able to predict the T4 lysozome (T4L)-benzene binding kinetic rates of ($k_{on}$, $k_{off}$) at ($0.53\pm0.08x10^7M^{-1}s^{-1}$, $791\pm197s^{-1}$), being highly consistent with the corresponding experimental value of ($0.08\text{-}0.1x10^7M^{-1}s^{-1}$, $950\pm200s^{-1}$). MSM was able to simultaneously predict the ligand association and dissociation rates through longer aggregated cMD simulations. For example, one MSM built with 59 μs cMD simulation data was able to accurately predict T4L-benzene binding kinetic rates. The predicted binding kinetic rate values of ($k_{on}$, $k_{off}$) were ($0.21\pm0.09x10^7M^{-1}s^{-1}$, $310\pm130s^{-1}$), being highly consistent with the experimental data of (0.08-0.1 $x10^7M^{-1}s^{-1}$, $950\pm200$ $s^{-1}$). MSM built with 50 μs cMD simulation data was used to predict the binding kinetic rates of the trypsin-benzamidine system[29b]. The predicted values of ($k_{on}$, $k_{off}$) were ($15.0\pm2.0x10^7M^{-1}s^{-1}$, $9.5\pm3.310^4s^{-1}$), being in line with the experimental values of ($2.9x10^7M^{-1}s^{-1}$, $600s^{-1}$). However, these calculations required very expensive computational resources.

Metadynamics[39] has been widely applied to investigate the ligand binding kinetics. Multiple Infrequent Metadynamics (InMetaD) simulations with a total of 5 μs trajectories were performed to predict the pathways of benzamidine binding to the trypsin and the binding kinetic rates. The predicted values of ($k_{on}$, $k_{off}$) were ($1.18\pm1.0$ x$10^7 M^{-1}s^{-1}$, $9.1 \pm 2.5$ $s^{-1}$), being smaller than the experimental values of ($2.9$x$10^7 M^{-1}s^{-1}$, $600 s^{-1}$). Similar smaller predicted values of ($k_{on}$, $k_{off}$) at ($0.0035\pm0.002$ x$10^7$ $M^{-1}s^{-1}$, $7\pm2$ $s^{-1}$) were observed in another 12 μs InMetaD simulations of benzene binding to T4L[28a]. For the Src-Dasatinib system, one study with 7 μs InMetaD simulations[40] was able to predict the $k_{off}$ value of $0.048\pm0.024$ $s^{-1}$, being highly consistent with the experimental value of 0.06 $s^{-1}$. For the p38α-compound I system, 6.8 μs InMetaD simulations[28f] predicted the $k_{off}$ value of $0.020 \pm 0.011$ $s^{-1}$, being in line with the experimental value of 0.14 $s^{-1}$. Besides, accuracy of force field also plays a critical role in predicting biomolecular binding kinetic rates. For example, Capelli et al.[41] applied two approaches to obtain the RESP charges for drug Iperoxo to predict its dissociation rate in the M2 receptor. The two approaches included the one with Amber standard methodology based on HF/6-31G* (RESP-HF) calculations and another one based on DFT/B3LYP (RESP-B3LYP) calculations. The simulations based on RESP-HF charges failed to predict the $k_{off}$ rate due to the unreasonable obtained transition state free energy. While simulations with RESP-B3LYP charges could predict the $k_{off}$ value of $3.7 \pm 0.7 \times10^{-4}$ $s^{-1}$, being in line with the experimental data of $1.0 \pm 0.2\times10^{-2}$ $s^{-1}$. For the Src-Imatinib system, Haldar et al.[42] showed that accounting for changes in charge distribution with QM/MM calculations improved the Imatinib dissociation rate from $0.0114 s^{-1}$ to $0.026 s^{-1}$, being more consistent to the experimental value of $0.11\pm0.08$ $s^{-1}$. Although Metadynamics simulations have shown remarkable improvements in capturing ligand binding and dissociation processes that occur over exceedingly long timescales, users often face a challenge for defining collective variables (CVs), which requires expert knowledge of the studied systems[43]. The simulations may suffer from a "hidden energy barrier" problem if important CVs were missed during the simulation setup[44]. To facilitate the choice of CVs, machine learning (ML) has been incorporated into Metadynamics simulations.

Wang et al. developed a predictive information bottleneck (PIB) approach to identify CVs and predict biomolecular dissociation rates [45]. The PIB was tested on the system of benzene binding to T4L and the predicted $k_{off}$ value was $3.3 \pm 0.8 s^{-1}$, being consistent with other InMetaD simulations but needing much shorter simulations[29c]. In another study, Filizola et al.[46] developed a novel approach, which combined InMetaD and ML methods including automatic mutual information noise omission and reweighted autoencoded variational Bayes to predict the dissociation kinetic rates of two drugs (morphine and bruprenorphine) in the μ-opioid receptor. Based on ~6 μs InMetaD simulations, the predicted $k_{off}$ for the morphine and bruprenorphine were $0.057 \pm 0.005\ s^{-1}$ and $0.021 \pm 0.003\ s^{-1}$, respectively, being within one order of magnitude difference from experimental values of $0.0023 \pm 0.001\ s^{-1}$ and $0.0018 \pm 0.03\ s^{-1}$. Very recently, Narjes et al.[47] combined ML and a novel Metadynamics approach, On-the-fly Probability Enhanced Sampling (OPES) flooding, to investigate the binding of benzamidine to trypsin. Based on a total of ~2.74 μs OPES simulations, they captured 55 benzamidine unbinding events and predicted the $k_{off}$ value of 1560 $s^{-1}$, being highly consistent with the experimental data.

Scaled MD[48] has been mainly used for the prediction of $k_{off}$ as a scale factor ranging from 0 to 1 is introduced in the simulations to reduce energy barrier to facilitate ligand dissociation. For example, Schuetz et al.[31b] performed scaled MD simulations to accurately predict the residence time and drug dissociation pathways of different inhibitors in Hsp90. In a recent study[49], Bianciotto et al. applied scaled MD simulations to predict the residence time and ligand unbinding pathways for a set of 27 ligands of Hsp90, being highly consistent with experimental data. In the RAMD simulations, an additional random force is applied on the ligand to promote its movement. Similar to scaled MD, RAMD is mainly used in the ligand dissociation simulations to qualitatively predict dissociation rates. In one recent study, Nunes-Alves et al.[30a] performed RAMD simulations to predict ligand dissociation rates of T4L. The predicted kinetic rates correlated well with experimental values for various systems with different ligands, temperatures, and protein mutations.

GaMD[27] is developed to apply a harmonic boost potential to enhance sampling with reduced energetic noise. The boost potential normally exhibits a near Gaussian distribution, which enables proper reweighting of the free energy profiles through cumulant expansion to the second order[27]. GaMD has been successfully applied to simulate important biomolecular processes, including protein/RNA folding[27b, 50], ligand/protein/RNA binding[50a, 51], and protein conformational changes[51f, 52]. However, it remained challenging to accurately predict ligand binding kinetic rates through normal GaMD [27a, 53]. Recently, a "selective GaMD" algorithm, called Ligand GaMD (LiGaMD) [54], has been developed to allow for more efficiently sampling of ligand binding and dissociation processes, which thus allows to accurately predict the ligand binding kinetic rates. For the protein ligand binding system, the system contains ligand L, protein *P* and the biological environment *E*. The system potential energy could be decomposed into the following terms:

$$V(r) = V_{P,b}(r_P) + V_{L,b}(r_L) + V_{E,b}(r_E) + V_{PP,nb}(r_P) + V_{LL,nb}(r_L) + V_{EE,nb}(r_E) + V_{PL,nb}(r_{PL}) + V_{PE,nb}(r_{PE}) + V_{LE,nb}(r_{LE}), \quad (6)$$

where $V_{P,b}$, $V_{L,b}$ and $V_{E,b}$ are the bonded potential energies in protein *P*, ligand $L$ and environment *E*, respectively. $V_{PP,nb}$, $V_{LL,nb}$ and $V_{EE,nb}$ are the self non-bonded potential energies in protein *P*, ligand $L$ and environment $E$, respectively. $V_{PL,nb}$ , $V_{PE,nb}$ and $V_{LE,nb}$ are the non-bonded interaction energies between *P-L*, *P-E* and *L-E*, respectively. Ligand binding mainly involves the non-bonded interaction energies of the ligand. Therefore, LiGaMD selectively boosts on the ligand essential energy term of $V_{ligand}(r) = V_{LL,nb}(r_L) + V_{PL,nb}(r_{PL}) + V_{LE,nb}(r_{LE})$ . In order to facilitate ligand rebinding, another boost was added to the remaining potential interaction of the system. Repetitive binding and dissociation of small-molecule ligands were captured in the LiGaMD simulations of host-guest and protein-ligand binding model systems[54]. Repetitive guest binding and dissociation in the β-cyclodextrin host were observed in hundreds-of-nanoseconds LiGaMD simulations. Accelerations of ligand kinetic rates in LiGaMD simulations were properly estimated using Kramers' rate theory. Furthermore, microsecond LiGaMD simulations observed repetitive benzamidine binding and dissociation in trypsin. The benzamidine binding and dissociation rates were predicted to be $1.15\pm0.79 \times 10^7$ $M^{-1}\cdot s^{-1}$ and $3.53\pm1.41$ $s^{-1}$, respectively. These data were comparable to the experimental values[55] of $2.9 \times 10^7$ $M^{-1}\cdot s^{-1}$ and 600 $s^{-1}$. Very recently, five replicas of 5 μs LiGaMD simulations successfully captured repetitive Nirmatrelvir drug binding and dissociation in the 3CLpro binding domain[56]. The Nirmatrelvir binding and dissociation rates were predicted to be $3.20\pm0.21\times10^5$ $M^{-1}\cdot s^{-1}$ and $2.92\pm0.37\times10^3 s^{-1}$, respectively. As no available experimentally determined binding kinetic rates, the authors predicted the dissociation constant ($k_D$) from the predicted binding kinetic rates by equation $k_D = k_{off}/k_{on}$. Notably, the predicted $k_D$ was 9.10±0.29 nM, being highly consistent with the available experimental value of 7±3 nM[57], demonstrating high accuracy of the predicted binding kinetic rates from LiGaMD simulations.

**Protein-peptide binding kinetics**

In comparison with the extensively studied protein-small molecule binding, protein-peptide binding studies are much less although increasing number of peptide-based drugs are being licensed to market in recent years[58]. Large conformational changes of peptides often occur during binding to target proteins, bringing huge challenges for modeling[22b, 59]. For example, coupled folding-upon binding mechanism has been observed in serval systems of peptide binding to proteins[22b, 59]. Only few number of computational approaches have been implemented to predict peptide binding kinetic rates, including the InMetaD[59], Weighted Ensemble[60], MSM[61], and Peptide GaMD (Pep-GaMD)[62] (**Table 3**).

InMetaD simulations with three CVs have successfully predicted the peptide binding and dissociation rates for the system of p53-MDM2[59]. Based on 27 μs InMetaD simulations[59], the predicted values of ($k_{on}$, $k_{off}$) were ($0.43 \pm 0.22 x 10^7 M^{-1}s^{-1}$, $0.7 \pm 0.4 s^{-1}$), being comparable to the corresponding experimental values of ($0.92x10^7 M^{-1}s^{-1}$, $2.06 s^{-1}$). Weighted Ensemble of a total amount of ~120 μs cMD simulations in implicit solvent was performed on the same p53-MDM2 system[60]. The predicted p53 binding kinetic rate ($k_{on}$) was $7 s^{-1}$, being highly consistent with the experiential data of $2.06\ s^{-1}$. Built on a total of 831 μs cMD simulations of p53 binding to the MDM2, the MSM[61] predicted accurate values of $k_{on}$ and $k_{off}$ at $0.019x10^7\ M^{-1}s^{-1}$ and $2.5\ s^{-1}$, respectively. However, the simulations needed for building MSM are much longer than the Weighted Ensemble and InMetaD simulations. Another MSM built on hundreds-of-microsecond cMD and Hamiltonian replica exchange simulations has been implemented to characterize binding and dissociation of the PMI peptide to the MDM2[63]. The PMI dissociation process is rather slow with the residence time at the timescale of second. Therefore, ~50 μs Hamiltonian replica exchange simulations were performed to predict the dissociate rate. The predicted values of ($k_{on}$, $k_{off}$) were ($300x10^7 M^{-1}s^{-1}$, $0.125\text{-}1.13 s^{-1}$), being comparable to the corresponding experimental values of ($52.7x10^7 M^{-1}s^{-1}$, $0.037 s^{-1}$).

Based on GaMD, we recently developed an algorithm called peptide GaMD or "Pep-GaMD" that enhances sampling of protein-peptide interactions[62]. As above mentioned, large conformational change involved in the process of peptide binding to target proteins[22b, 59]. Therefore, peptide binding involves in both the bonded and non-bonded interaction energies of the peptide. Thus, the essential peptide potential energy is defined as $V_{peptide}(r) = V_{LL,b}(r_L) + V_{LL,nb}(r_L) + V_{PL,nb}(r_{PL}) + V_{LE,nb}(r_{LE})$. A selective boost was thus added to the essential peptide potential to facilitate the dissociation of peptides in the Pep-GaMD. In addition to selectively boosting the peptide, another boost potential is applied on the protein and solvent to enhance conformational sampling of the protein and facilitate peptide rebinding.

Pep-GaMD[62] has been developed to capture repetitive peptide binding and dissociation processes, which allows us to calculate the peptide binding free energies and kinetic rates. It has been demonstrated on binding of three model peptides to the SH3 domains[64], including "PPPVPPRR" (PDB: 1CKB), "PPPALPPKK" (PDB: 1CKA) and "PAMPAR" (PDB: 1SSH). Repetitive peptide binding and unbinding events were captured in independent 1 μs Pep-GaMD simulations, allowing us to calculate peptide binding thermodynamics and kinetics. The predicted values of both binding free energies and kinetic rates from Pep-GaMD simulations were in good agreement with available experimental data. Particularly, the predicted peptide binding kinetic rates of 1CKB was ($4060 \pm 2260 \times 10^7$ $M^{-1} \cdot s^{-1}$, $1450 \pm 1170$ $s^{-1}$), being within 1 order of the experimental data of ($150 \times 10^7$ $M^{-1} \cdot s^{-1}$, 8900 $s^{-1}$).

**Protein-protein binding kinetics**

Protein-protein interactions (PPIs) play key roles in many fundamental biological processes, including cellular signal transduction, immune responses and so on[1]. Moreover, PPIs are implicated in the development of numerous human diseases and served as important drug targets.[65] PPIs exhibit unique features, being distinct from the protein-small molecule and protein-peptide interactions. The protein-protein binding affinity is often stronger than that of protein-small molecule and protein-peptide interactions. Protein-protein binding and unbinding processes often occurred in significantly longer timescale. Particularly, protein-protein dissociation process could take place in a much longer time scale, from seconds to even days. Tens of microseconds cMD simulations were able to capture barnase binding to barstar[20d]. Based on 28 successfully binding events captured in a total of ~213 μs Anton cMD simulations with TIP4P2005 water model[20d], the predicted barnase binding rate ($k_{on}$) was $6x10^7 M^{-1}s^{-1}$, being in line with the experimental value of $60x10^7 M^{-1}s^{-1}$. Less barnase binding events (24) with slower predicted binding rate ($2.3x10^7 M^{-1}s^{-1}$) were observed with the TIP3P water model. Additionally, Pan et al.[20d] successfully predicted the binding kinetic association rates of another two systems of insulin dimerization and Ras binding to Ras-binding domain of c-RAF-1 (Ras-Raf-RBD). Based on 6 successful binding events among the total of 294.8 μs cMD simulations, the predicted association rate ($k_{on}$) of the insulin dimerization was $0.41x10^7 M^{-1} \cdot s^{-1}$, being comparable to the experimental value of $11.4x10^7 M^{-1} \cdot s^{-1}$. For the Ras-Raf-RBD system, 117 μs cMD simulations successfully captured 7 binding events and predicted $k_{on}$ value of $2.6 x10^7 M^{-1} \cdot s^{-1}$, being highly consistent with the experimental data of $4.5 x10^7 M^{-1} \cdot s^{-1}$. However, it remains challenging to simulate the protein dissociation with cMD[20d].

Weighted Ensemble[24d] of a total of ~18 μs cMD simulations were able to capture 203 barnase binding events and accurately predict the barnase-barstar binding rate constant ($k_{on}$) of $23 \pm 10 x10^7 M^{-1} \cdot s^{-1}$. Plattner et al.[66] performed high throughput MD simulations of the barnase binding to barstar to build MSM. A total of 1700 μs cMD simulations with 1,892 independent replicas starting from unbound state captured 74 barnase binding events. Another set of 300 μs adaptive MD simulations captured 16 and 10 times of barnase binding and dissociation events, respectively. Based on the total of 2,000 μs simulation data, the obtained MSM was able to predict intermediate structures, binding energies and kinetic rates that were consistent with experimental data[66].

Recently, we developed a selective PPI-GaMD method[67] to simulate repetitive protein binding and dissociation in order to calculate protein binding free energies and kinetics. The PPI simulation system consists of a ligand protein L, a target protein P and a biological environment E. In PPI-GaMD, a selective boost potential is added to the non-bonded protein-protein interaction energy $V_{PL,nb}$. Another boost potential is applied on the remaining potential energy of the system to enhance conformational sampling of the proteins and facilitate protein diffusion and rebinding[67]. PPI-GaMD[62] has been demonstrated on the model system of barnase binding to the barstar. Six independent 2 μs PPI-GaMD simulations have successfully captured repetitive barstar dissociation and rebinding events. Three to six binding and dissociation events were observed in each individual PPI-GaMD simulations. The barnase binding free energy predicted from PPI-GaMD was -17.79 kcal/mol with a standard deviation of 1.11 kcal/mol, being highly consistent with the experimental value of -18.90 kcal/mol[65c]. Additionally, the PPI-GaMD simulations allowed us to calculate the protein binding kinetics. The average $k_{on}$ and $k_{off}$ were predicted as $21.7\pm13.8\times10^{8}$ $M^{-1}\cdot s^{-1}$ and $7.32\pm4.95\times10^{-6}$ $s^{-1}$, being consistent with the corresponding experimental values of $6.0\times10^{8}$ $M^{-1}\cdot s^{-1}$ and $8.0\times10^{-6}$ $s^{-1}$, respectively.

## 5. Conclusions and outlook

Both experimental and computational techniques have achieved remarkable advances in characterizing biomolecular binding kinetics, including SPR, QSKR, MD and enhanced sampling simulations. It is still very expensive and resource-consuming for experimental techniques to obtain biomolecular binding kinetic rates. Nevertheless, recent years have seen increasing numbers of experimental binding kinetic data, leading to a number of databases to collect such information.

Based on the experimental binding kinetic data, QSKRs have been developed to predict binding kinetic rate constants with high throughtput[15]. For MD simulations, accuracy of binding free energy calculations could be within 1.0 kcal/mol with the modern techniques[68]. Compared with extensively studied biomolecular binding thermodynamics, the accuracy and efficacy of modeling techniques for predicting biomolecular binding kinetics are still not very high. The predicted binding kinetic rate constants from MD simulations and related enhanced sampling methods could derivate orders of magnitude from the experimental data (**Tables 2-4 & Fig 1B**). Nevertheless, MD simulations have enabled characterization of biomolecular binding pathways and kinetics, attracting increasing attentions in recent years. With advances in computer hardware and accuracy of force fields, long timescale cMD simulations with all-atom and/or coarse-grained models have successfully captured biomolecular binding process and predicted accurate binding associate rates[20d], although slower dissociation processes are still difficult to simulate.

Enhanced sampling methods have greatly reduced the computational cost for calculations of biomolecular kinetics. Among various enhanced sampling methods, the MSM, InMetady and GaMD appear to be the most used techniques that allow for simultaneous predictions of biomolecular binding association and dissociation rates (**Fig. 2**). Another trend is the incorporation of ML into enhanced sampling methods to further improve sampling efficiency and prediction accuracy of biomolecular binding kinetic rates[46, 69].

Overall, current computational methods have been tested mostly on model systems with published experimental kinetic data in the literature. The simulation protocols could be potentially calibrated to predict the kinetic rate constants against the experimental values. This would suggest a need for community blind challenges on biomolecular binding kinetics predictions, in which participants predict the kinetic rates without knowing the experimental values and the predictions will be evaluated independently by the challenge organizers. Such challenges are expected to greatly facilitate improvements of the various techniques developed for predicting biomolecular binding kinetics in the field. In addition to protein-ligand binding, protein-peptide binding and protein-protein interactions, interactions of nucleic acids (RNA and DNA) with small molecules and proteins remain largely underexplored and warrant more kinetics studies.

In summary, accurate calculations of biomolecular binding kinetics of large biomolecular complexes present grand challenges for computational modelling and enhanced sampling simulations. Further innovations in both computing hardware and method developments may help us to address these challenges in the future.

**Acknowledgements**

This work used supercomputing resources with allocation awards TG-MCB180049 and BIO210039 through the Extreme Science and Engineering Discovery Environment (XSEDE), which is supported by National Science Foundation grant number ACI-1548562 and project M2874 through the National Energy Research Scientific Computing Center (NERSC), which is a U.S. Department of Energy Office of Science User Facility operated under Contract No. DE-AC02-05CH11231. It also used computational resources provided by the Research Computing Cluster at the University of Kansas. This work was supported in part by the National Institutes of Health (R01GM132572) and National Science Foundation (2121063).

**Table 1** Databases of biomolecular binding kinetics.

| Database | Description | Website |
|---|---|---|
| KDBI | It includes 19,263 entries, which provides experimentally verified kinetic rates for protein-protein/DNA/RNA/ligand and ligand-DNA/RNA interactions. | http://xin.cz3.nus.edu.sg/group/kdbi/kdbi.asp |
| BindingDB | It focuses on protein-ligand interaction, including ~1.1 million compounds and 8.9 thousand targets. | https://www.bindingdb.org/rwd/bind/index.jsp The webpage of binding kinetic rates: https://bindingdb.org/rwd/bind/ByKI.jsp?specified=Kn |
| KOFFI | It includes 1705 entries and a rating system to measure the quality of experimental data. | http://koffidb.org/ |
| PDBbind | The $k_{off}$ dataset includes 680 entries with protein-small molecule complex structure. | http://www.pdbbind.org.cn/ |
| SKEMPI | It focuses on protein-protein interaction, which records 713 binding association and dissociation rates upon mutation. | http://life.bsc.es/pid/mutation_database/ |
| dbMPIKT | It focuses on protein-protein interaction, which contains 5291 protein binding association and dissociation rates upon mutation. | http://deeplearner.ahu.edu.cn/web/dbMPIKT/ |

**Table 2** Summary of computer simulation predicted protein-ligand binding ($k_{on}^{sim}$) and dissociation ($k_{off}^{sim}$) rates compared with experimentally determined binding ($k_{on}^{exp}$) and dissociation ($k_{off}^{exp}$) rates.

| System | Method | $k_{on}^{exp}$ ($10^7 M^{-1}s^{-1}$) | $k_{off}^{exp}$ ($s^{-1}$) | $k_{on}^{sim}$ ($10^7 M^{-1}s^{-1}$) | $k_{off}^{sim}$ ($s^{-1}$) | Sim. time (μs) | $\Delta log k_{on}$ | $\Delta log k_{off}$ | Force field | Year[Ref] |
|---|---|---|---|---|---|---|---|---|---|---|
| Trypsin-Benzamidine | M-WEM | 2.9 | 600 | 0.53±0.08 | 791±197 | 0.48 | -0.74 | 0.12 | AMBER ff14SB and GAFF | 2022[38] |
| Trypsin-Benzamidine | SEEKR2 | 2.9 | 600 | 2.4±0.2 | 990±130 | 5 | -0.082 | 0.22 | - | 2022[70] |
| Trypsin-Benzamidine | InMetaD+ML | 2.9 | 600 | - | 1560 | 2.75 | - | 0.41 | AMBER ff14SB force field and GAFF | 2022[47] |
| Abl kinase-imatinib | Mile-stoning | 0.15±0.01 | 25±6 | -- | 18 | 1.043 | - | -0.14 | CHARMM 36 and CGenFF | 2021[26] |
| Trypsin-Benzamidine | LiGaMD | 2.9 | 600 | 1.15±0.79 | 3.53±1.41 | 5 | -0.40 | -2.23 | AMBER ff14SB and GAFF | 2020[54] |
| Trypsin-Benzamidine | SEEKR | 2.9 | 600 | 12±0.5 | 174±9 | 4.4 | 0.62 | -0.54 | - | 2020[71] |
| M2-Iperoxo | Frequency-adaptive MetaD | - | 0.01±0.002 | - | $3.7 \pm .0.7 \times 10^{-4}$ | 8 | - | -1.43 | AMBER14SB and GAFF | 2020[41] |
| Trypsin-Benzamidine | CGMD | 2.9 | 600 | 36.8 | $6.9 \times 10^{5}$ | 428 | 1.10 | 3.06 | MARTINI | 2020[34] |
| μOR-morphine | InMetaD+ML | 0.29±0.001 | 0.023±0.001 | - | 0.057±0.005 | 6 | - | 0.39 | CHARMM 36 and CGenFF | 2020[46] |
| μOR-bruprenorphine | InMetaD+ML | 1.33±0.01 | 0.0018±0.003 | - | 0.021±0.003 | 19 | - | 1.07 | CHARMM 36 and CGenFF | 2020[46] |
| Src-Dasatinib | cMD | 0.5 | 0.06 | 0.76 | - | 6.6 | 0.18 | - | OPLS | 2020[32] |
| Src-Dasatinib | CGMD | 0.5 | 0.06 | 4 | - | 300 | 0.90 | - | MARTINI | 2020[33b] |
| Trypsin-Benzamidine | InMetaD | 2.9 | 600 | - | 4176±324 | ~1.00 | - | 0.84 | CHARMM 36 and CGenFF | 2019[72] |
| Trypsin-Benzamidine | WE | 2.9 | 600 | - | 2660 | 8.75 | - | 0.65 | CHARMM and CGenFF | 2019[37] |
| T4L-BEN | ML | 0.08-0.1 | 950±200 | - | 3.3±0.8 | - | - | -2.46 | CHARMM22* | 2019[45] |
| T4L-BEN | InMetaD | 0.08-0.1 | 950±200 | 0.0035±0.002 | 7±2 | 12 | -1.36 | -2.13 | CHARMM22* and CGenFF | 2018[28a] |
| T4L-BEN | MSM | 0.08-0.1 | 950±200 | 0.21±0.09 | 310±130 | 59 | 0.42 | -0.49 | CHARMM36 | 2018[29c] |
| T4L-BEN | WE | 0.08-0.1 | 950±200 | - | 1000 | 29 | | 0.022 | CHARMM36 | 2018[25b] |
| Src-Imatinib | MetaD | - | 0.11±0.08 | - | 0.026 | - | - | -0.63 | AMBER FF99SB-ILDN and GAFF | 2018[42] |
| Src-Dasatinib | InMetaD | 0.5 | 0.06 | - | 0.048±0.024 | 7 | - | -0.096 | OPLS | 2017[40] |
| P38α-compound I | InMetaD | 0.0118 | 0.14 | - | 0.02±0.01 | 6.8 | - | -0.84 | AMBER FF99SB-ILDN and GAFF | 2017[28f] |
| Trypsin-Benzamidine | InMetaD | 2.9 | 600 | 1.18±1.0 | 9.1±2.5 | - | -0.39 | -1.82 | AMBER ff99SB-ILDN and GAFF | 2015[73] |
| Trypsin-Benzamidine | MSM | 2.9 | 600 | 15±2 | $9.5 \pm 3.3 \times 10^{4}$ | 50 | 0.71 | 2.20 | AMBER ff99SB and GAFF | 2011[29a] |
| Src-Dasatinib | cMD | 0.5 | 0.06 | 0.19 | - | 35 | -0.42 | - | AMBER ff99SB and GAFF | 2011[22a] |

**Table 3** Summary of computer simulation predicted peptide binding ($k_{on}^{sim}$) and dissociation ($k_{off}^{sim}$) rates compared with experimentally determined binding ($k_{on}^{exp}$) and dissociation ($k_{off}^{exp}$) rates.

| System | Method | $k_{on}^{exp}$ ($10^7 M^{-1} s^{-1}$) | $k_{off}^{exp}$ ($s^{-1}$) | $k_{on}^{sim}$ ($10^7 M^{-1} s^{-1}$) | $k_{off}^{sim}$ ($s^{-1}$) | Sim. time (μs) | $\Delta log k_{on}$ | $\Delta log k_{off}$ | Force field | Year[Ref] |
|---|---|---|---|---|---|---|---|---|---|---|
| SH3-1CKB | Pep-GaMD | 150 | 8900 | 4060±2260 | 1450±1170 | 3 | 1.43 | -0.79 | AMBER ff14SB | 2020[62] |
| MDM2/P53 | InMetaD | 0.92 | 2.06 | 0.43±0.22 | 0.7±0.4 | 27 | 0.88 | -0.47 | AMBER ff99SB-ILDN | 2020[59] |
| MDM2/PMI | MSM | 52.7 | 0.037 | 330 | 0.125-1.13 | 500 | 0.80 | 0.53 | AMBER ff99SB-ILDN | 2017[63] |
| MDM2/P53 | MSM | 0.92 | 2.06 | 0.019 | 2.5 | 831 | 0.88 | 0.08 | AMBER ff99SB-ILDN-nmr | 2017[61] |
| MDM2/P53 | WE | 0.92 | 2.06 | 7 | - | 120 | 0.88 | - | AMBER ff99SB-ILDN | 2016[60] |

**Table 4.** Summary of computer simulation predicted protein-protein binding ( $k_{on}^{sim}$ ) and dissociation ($k_{off}^{sim}$) rates compared with experimentally determined binding ($k_{on}^{exp}$) and dissociation ($k_{off}^{exp}$) rates.

| System | Method | $k_{on}^{exp}$ ($10^7 M^{-1}s^{-1}$) | $k_{off}^{exp}$ ($s^{-1}$) | $k_{on}^{sim}$ ($10^7 M^{-1}s^{-1}$) | $k_{off}^{sim}$ ($s^{-1}$) | Sim. time (μs) | $\Delta log k_{on}$ | $\Delta log k_{off}$ | Force field | Year[Ref] |
|---|---|---|---|---|---|---|---|---|---|---|
| Barnase-Barstar | PPI-GaMD | 60 | $8x10^{-6}$ | 217±138 | $7.32±4.95x10^{-6}$ | 12 | 0.56 | -0.038 | AMBER ff14SB | 2022[67] |
| Barnase-Barstar | WE | 60 | $8x10^{-6}$ | 230±100 | - | 18 | 0.58 | - | AMBER ff03* | 2019[24d] |
| Barnase-Barstar | cMD | 60 | $8x10^{-6}$ | 2.3 | - | 440 | -1.42 | - | AMBER ff99SB-ILDN | 2019[20d] |
| Insulin Dimer | cMD | 11.4 | 14800 | 0.41 | - | 294.8 | -1.44 | - | AMBER ff99SB-ILDN | 2019[20d] |
| Ras–Raf-RBD | cMD | 4.5 | 7.4 | 2.6 | - | 117 | -0.24 | - | AMBER ff99SB-ILDN | 2019[20d] |
| Barnase-Barstar | MSM | 60 | $8x10^{-6}$ | 26.3-26.5 | $3x10^{-6}$ | 1700 | -0.36 | -0.42 | AMBER ff99SB | 2017[66] |

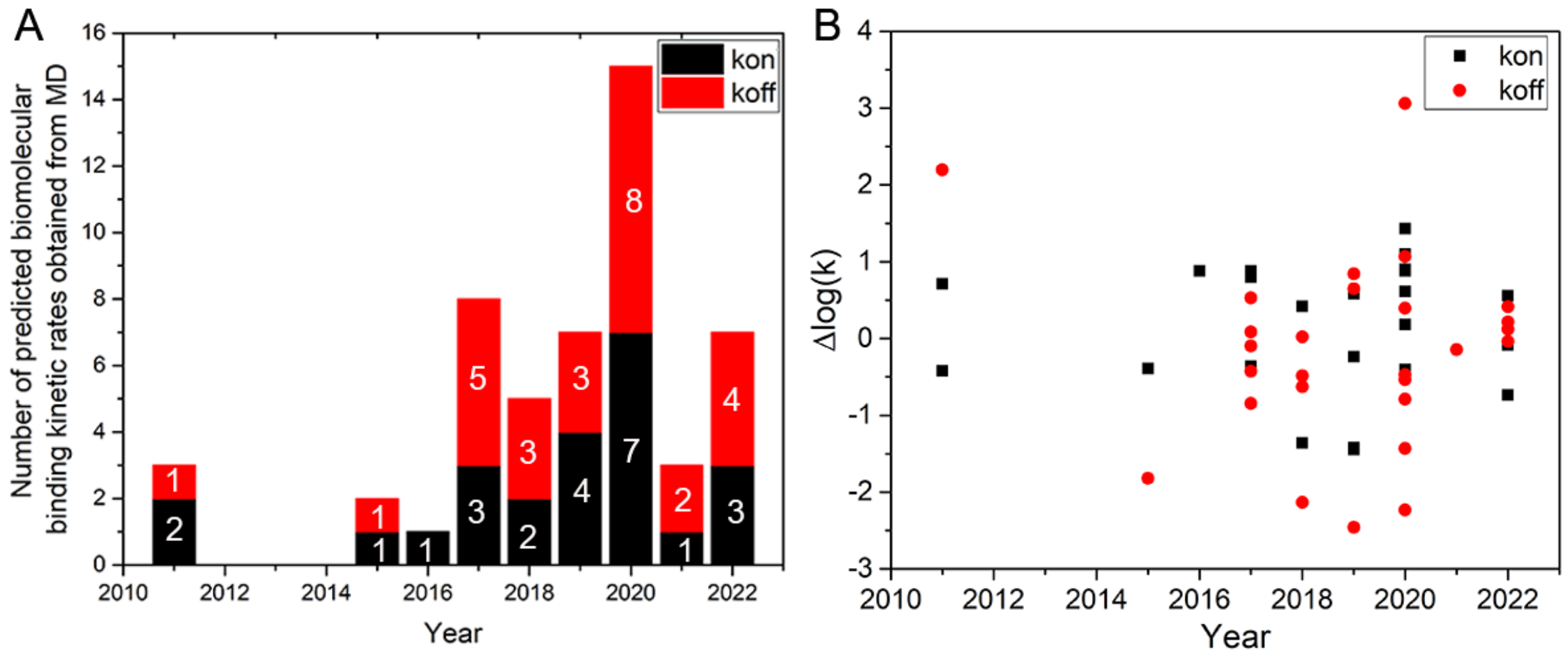

**Figure 1**. The number (**A**) and accuracy (**B**) of predictions of biomolecular binding kinetic rates obtained from MD simulations plotted over the years.

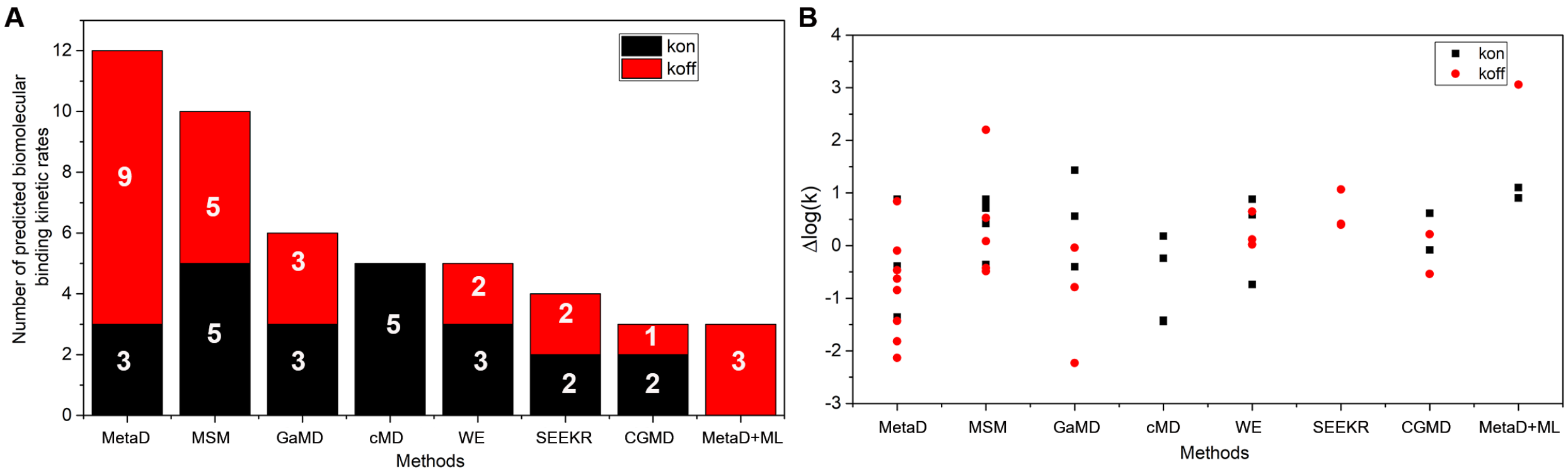

**Figure 2.** The number (**A**) and accuracy (B) of predicted biomolecular binding kinetic rates using different MD techniques, including Metadynamics (MetaD), Markov State Models (MSM), Gaussian accelerated MD (GaMD), conventional MD (cMD), Weighted Ensemble (WE), simulation enabled estimation of kinetic rates (SEEKR), coarse-grained MD (CGMD) and combination of Metadynamics and Machine Learning (MetaD+ML).